# Reducing Honeypot Log Storage Capacity Consumption – Cron Job with Perl-Script Approach

## Iman Hazwam Abd Halim[1]*, Nur Muhammad Irfan Abu Hassan[2], Tajul Rosli Razak[3], Muhammad Nabil Fikri Jamaluddin[4], Mohammad Hafiz Ismail[5]

[1,2,3,4,5]Faculty of Computer and Mathematical Sciences,
*Universiti Teknologi MARA, 02600 Arau, Perlis, MALAYSIA*

*Corresponding author: *hazwam688@uitm.edu.my*

## ABSTRACT

*Honeypot is a decoy computer system that is used to attract and monitor hackers' activities in the network. The honeypot aims to collect information from the hackers in order to create a more secure system. However, the log file generated by honeypot can grow very large when heavy traffic occurred in the system, such as Distributed Denial of Services' (DDoS) attack. The DDoS possesses difficulty when it is being processed and analyzed by the network administrator as it required a lot of time and resources. Therefore, in this paper, we propose an approach to decrease the log size that is by using a Cron job that will run with a Perl-script. This approach parses the collected data into the database periodically to decrease the log size. Three DDoS attack cases were conducted in this study to show the increasing of the log size by sending a different amount of packet per second for 8 hours in each case. The results have shown that by utilizing the Cron job with Perl-script, the log size has been significantly reduced, the disk space used in the system has also decreased. Consequently, this approach capable of speeding up the process of parsing the log file into the database and thus, improving the overall system performance. This study contributes to providing a pathway in reducing honeypot log storage using the Cron job with Perl-Script.*

16

# INTRODUCTION

Nowadays, the world is increasingly relying on computer networks. The use of network resources is increasing followed by many risks toward security problems. There are many vulnerabilities and threats that were discovered every day and are affecting users and companies at critical stages, from privacy issues to financial losses. Recently, many organization including government agencies have been attacked and scanned by unknown intruders which resulting in financial and reputation damages (Popereshnyak, Suprun, Suprun, & Wieckowski, 2018). Monitoring network activity is a must for network administrators and security analysts to understand these threats and to develop a secure network environment (Darapareddy & Gummadi, 2012).

Intrusion Detection System (IDS) is one of the solutions that can be used to monitor network traffic. It is a software application that can identify any suspicious activities and issue alerts to network administrator when such activity is identified. To ensure the network can be secured, honeypot-based intrusion detection system was introduced to monitor unused Internet Protocol (IP) spaces to learn about attackers. The advantage of honeypots over other monitoring solutions is to collect only suspicious activities and obtain information about the attackers (Kondra, Janardhan Reddy Bharti, Santosh Kumar Mishra, Sambit Kumar Babu, 2016). Although an IDS does not have the capability to prevent malware or any type of attacks, the technology is still relevant to be used in the modern enterprise because of its functionality that is able to detect an active attack (Place, 2018).

One of the most widely used tools is honeyd for creating honeypots. Honeyd is an open source of the computer program created by Niels Provos that allows a user to set up and run multiple virtual hosts on a computer network (Provos, 2019). It is one of the low-interaction honeypot which can be used to simulate 65,000 hosts using a single PC. The IP stack of different OS can be emulated through honeyd. It is designed to resemble the real system and equipped with the already known vulnerability so that the attacker will be distracted from the main system that he will attack and moves to the false honeypot (Sembiring, 2016). In a network, all normal traffic should be forwarded to and from valid servers only. Thus, a network administrator running honeyd can monitor logs to see if there is any traffic going to the virtual hosts set up by honeyd. Any traffic going to these virtual servers can be considered highly suspicious. The network administrator can take preventative actions, possibly by blocking the suspicious IP address or by further monitoring the network for suspicious traffic.

The problem is that the logs generated by honeyd can grow very large when heavy traffic occurred in the system such as Distributed Denial of Services (DDoS) attack, which can consume a lot of disk space (Fahd & Saleh, 2012; Joshi & Kakkar, 2017). A large amount of log size contains difficulties when they are processed and analysed by network administrator or security analysts as they required a lot of time and resources (Singh & Joshi, 2011).

Therefore, this paper has put forward an approach to address these issues by introducing the '*cron job*' with perl-script approach to periodically transfers parsed log data into database. The approach is efficiently reduced usage of disk storage and system resources.

The rest of this paper is organised as follows. Section *Background* reviews the background of the intrusion detection system (IDS), distributed denial of services attack (DDoS Attack) and Cron job. This is followed by Section *Proposed Approach* that introduces the Cran job with perl-script as an approach, which consist of four key steps. Section *Experiment and Result* discusses the experimental works and results of the proposed approach. Finally, Section *Conclusion* presents the conclusion and future works.

## Background

### i. Intrusion Detection System (IDS)

Honeypot can be divided in two categories which is low interaction and high interaction. Both have its own strengths and weaknesses.

Low interaction honeypots simulate services in such a way that they cannot be exploited to gain complete access. In these types of honeypots, there is no operating system for the attackers to interact with. The deployment and maintenance process of low interaction honeypots is simple than high interaction honeypots. Low interaction honeypots can minimize risk but their functionality is limited. Honeyd and specter is an example of low interaction honeypot (Fahd & Saleh, 2012).

High interaction honeypots are considered the most advanced type of honeypots in general. It has its own operating system which allows the user to have no restrictions to perform whatever tasks and actions that are desired. Highest level of risk is associated with them as they are using an actual operating system. Designing the high interaction honeypots is a time consuming and the process can be difficult. However, it can collect a large amount of data as the attackers have most of the resources available to them, while all their actions and activities are logged on the honeypot (Eldad, 2018).

### ii. Distributed Denial of Services Attack (DDoS Attack)

The rapid development and popularization of internet had resulted in the increase of online attacks. When the attacker uses a single machine to interrupt the services, it is identified as Denial of Services (DoS) attack. The extended attack from DoS is DDoS which initiates the attack from multiple controlled devices. According to(Bhosale, Nenova, & Iliev, 2017), DDoS attack has arisen to be the most powerful and harmful attack. A malicious software called 'Bots' is injected into multiple computers to gain control to perform specific and automated function. Bots constructed in large number are also called as 'Botnet' have caused major crimes such as, Click-fraud, widespread delivery of Spam emails, spyware installation, worm and virus dissemination. Such attacks were able to gain access into the networks bandwidth and also resources of victims, thus increased the success rate of denial of access to legitimate users. The basic form of a DDoS attack is an online attack in which an attacker sends a large amount of traffic to a website or network. The purpose of DDoS attack is to overwhelm the server and network services until it crashes and fails to respond on incoming requests from legitimate users (Jessica, 2018).

### iii. Cron Job

Cron is a job scheduler program that enables Unix user to execute commands or shell scripts automatically at a fixed times, dates or intervals. When the jobs run, the user may not be present, and they are executed through a long-running daemon process belonging to the root user and then later will be transferred to the real user for further analysis (Dale, 2003).

## PROPOSED APPROACH

The proposed approach will be explained based on the model architecture as shown in Fig. 1. The model has been divided into four steps process which will include the installation and configuration procedure until the honeypot storage truncation process.

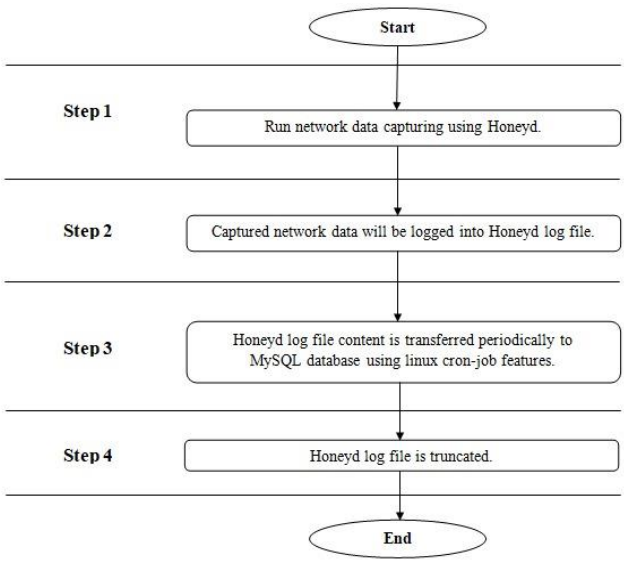

Fig. 1: Stepwise process in the proposed approach

### i. Step 1

The first step is to install and configure honeyd tool under Ubuntu 12.04 operating system platform. This device with honeyd will work as a honeypot which will capture all the network activity that has gone through the device. The LAMP (Linux, Apache, MySQL, PHP) web server are also required in creating the database to store all the network information captured from honeypot to another backup device or any cloud storage.

*ii. Step 2*

The next step of this approach; honeyd will done its job by capturing all the incoming data traffic from the clients. Honeyd will continuously monitor the unused IP space and whenever an attacker attempts to establish a connection and it acts as a victim. The captured network data then will be logged in the form of text file. This text files will be stored in the honeypot server.

*iii. Step 3*

By time, the honeypot server storage will become stuffed with the log file of the incoming data traffic especially when it encounters an unwanted network traffic situation such as DDOS attack. Thus, in order to sustain the operation of the honeypot, the honeyd log files need to be transferred periodically to other secondary database storage. In this case, this model has used MySQL as its database. Perl-script is used along with the cron job in order to parse the logged files into the database hourly.

*iv. Step 4*

After each of the log file transferring process completed, the log file in the honeypot storage will be truncated. During this process, capturing will halt for a while allowing data transfers to the database. The truncation is done in order to avoid a too large log file storage size usage in the honeypot that will affect the performance of the honeypot itself.

## EXPERIMENT AND RESULT

Several tests were done by using one client to perform DDoS attack in three different cases. Each attack was performed in the duration of 8 hours. Low Orbit Ion Canon (LOIC) tool was used to send a different number of TCP packet to the honeypot for each case in order to create DDOS attack.

*i. Case 1*

In the first case, the amount of 10 threads has produced the size of 85MB log file. The log file has consumed 1% of disk space in the system storage. It takes a total of 35 minutes to transfer content of log file into database. After using cron-job technique, it managed to shrink the file by 87% from 85MB into 11MB and reduced disk space used by 1%. This was achieved by deleting the old content of the log file after parsing process was done at every hour. When the log file capacity became less, the time taken to transfers data into database was reduced by 89% from 35 minutes to 4 minutes as it only required to transfers less amount of data.

### ii. Case 2

For the second case, it sent 50 threads per second which produced the size of 446MB log file. The log file consumed 6% of disk space in the system. In transferring the data from log file into database, it takes 3 hours and 10 minutes. After using cron-job technique, it managed to shrink the file into 56MB by 87% and reduced disk space used by 6%. This is achieved by deleting the old data in the log file after parsing process was done at every hour. As the log capacity was reduced, it only takes 23 minutes of transferring data into database. The time taken to parse the log file is reduced by 88% due to the less amount of data.

### iii. Case 3

In the third case, it sent 90 threads per second which produced the size of 844MB log file. The log file consumed 9% of disk space which lead to a total of 100% disk space usage in the system. At this point, the system performance started to drop down in terms of availability, response time, and the processing speed due to the full capacity in the system storage. It takes 5 hours and 20 minutes to transfer the data from log file into database which is time consuming and wasting system resources. After using cron-job technique, it managed to shrink the file into 118MB by 86% and reduced disk space used by 9%. This is achieved by deleting the old data in the log file after parsing process was done at every hour. The process of transferring data into database was reduced by 88% from 5 hours and 20 minutes to 40 minutes as less data is required to transfer.

### iv. Cron Job vs. Non-Cron Job

This section explains the comparison of the log file before and after using the Cran job in term of size, disk space and time, that discussed earlier. Hence, Table 1 presents all the findings that were obtained from testing and experiment in Cases 1, 2, and 3.

Table 1: Overall comparison before and after using Cron job

| Case | Size (MB) | Disk Space Used (%) | Time (minutes) | Cron-Job |
|---|---|---|---|---|
| 1 | 85 | 95 | 35 | x |
|   | 11 | 94 | 4 | √ |
| 2 | 446 | 91 | 190 | x |
|   | 56 | 85 | 23 | √ |
| 3 | 884 | 100 | 320 | x |
|   | 118 | 91 | 40 | √ |

In Fig. 2, the trend shows that the size of the log file was significantly decreased after using the Cron job from Case 1 to Case 3. The tremendous reduction can be seen in Case 3.

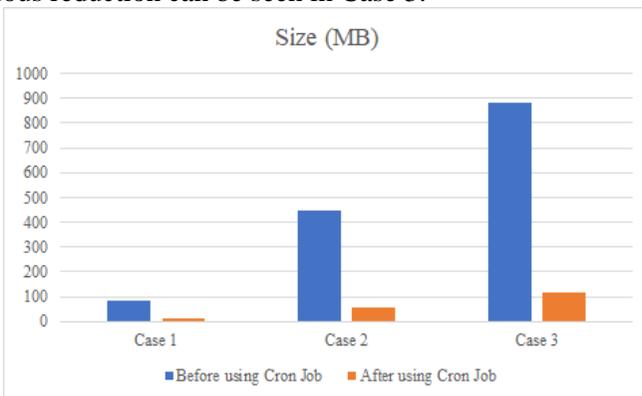

Fig. 2: Comparing size (MB) before and after using Cron job

Likewise, in the case of disk space of log file, the same pattern is observed that also shown the decrease after using the Cron job as shown in Fig. 3. However, this time, the reduction of disk space is small after using the Cron job.

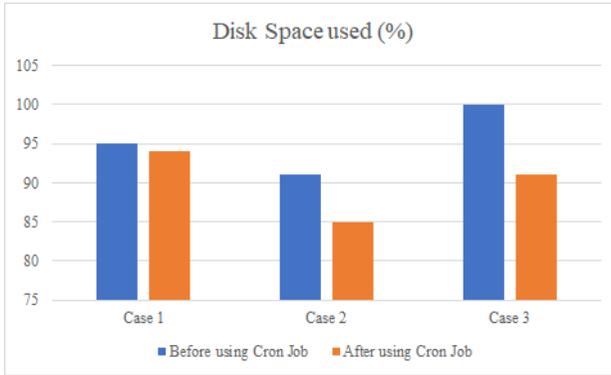

Fig. 3: Comparing disk space (%) before and after using Cron job

In the case of time taken to parse in the database, the significant reduction is recorded again after using the Cron job from Case 1 to Case 3. Also, a huge reduction can be viewed in Case 3.

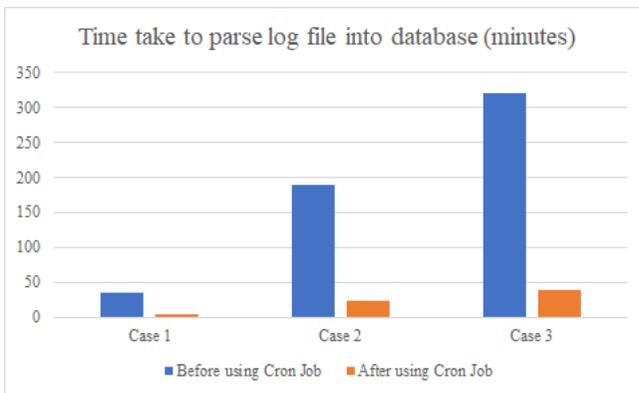

Fig. 4: Comparing time taken to parse log file into a database (%), before and after using Cron job

Overall, the Cron job approach managed to reduce the log file and disk space used, speed up the process of transferring data into the database, and able to mitigates flood attack. Obviously, other methods are available that also can be used; nevertheless, based on the current evidence, the Cron job approach shows promising toward improving overall system performance.

# CONCLUSION

Honeypot was usedto log and store data into the log file. In term of log size, the maximum size that was achieved is 844MB. There is still a possibility that the log size can be increase beyond the maximum size. However, the log file has stopped to grow due to the limited storage capacity of the system. If the system has more storage available, it canproduce a larger log size, thus the results obtained from the finding could be more accurate.

Currently, this approach only focuses on reducing the log size through parsing the data from the log file into the database. For the next step, other researchers may use different methods such as packet filtering. This method can be used to obtain only desired information during logging and able to overcome data redundancy, thus can also reduce the log size.Besides using cron job scheduling, it may also possibly be replaced by using log rotations approach where the old files will be archived, and the new log file is created.